\documentclass[galaxies,article,accept,oneauthors,dvi2pdf]{mdpi} 

\firstpage{1} 
\pubvolume{xx}
\issuenum{1}
\articlenumber{5}
\pubyear{2020}
\copyrightyear{2020}
\history{Received: date; Accepted: date; Published: date}
\Title{Constraints on Lorentz invariance violation with multiwavelength polarized astrophysical sources}

\Author{Qi-Qi Zhou$^{1}$, Shuang-Xi Yi$^{1}$*\orcidA, Jun-Jie Wei$^{2,3}$*\orcidB and Xue-Feng Wu$^{2,3}$*\orcidC}

\AuthorNames{Qi-Qi Zhou, Shuang-Xi Yi, Jun-Jie Wei and Xue-Feng Wu}

\address{%
  $^1$ \quad School of Physics and Physical Engineering, Qufu Normal University, Qufu 273165, China\\
  $^2$ \quad Purple Mountain Observatory, Chinese Academy of Sciences, Nanjing, 210023, China\\
  $^3$ \quad School of Astronomy and Space Sciences, University of Science and Technology of China, Hefei 230026, China}
\corres{Correspondence: yisx2015@qfnu.edu.cn; jjwei@pmo.ac.cn; xfwu@pmo.ac.cn}

\abstract{
Possible violations of Lorentz invariance (LIV) can produce vacuum birefringence, which results in a frequency-dependent rotation of the polarization plane of linearly polarized light from distant sources. In this paper, we try to search for a frequency-dependent change of the linear polarization angle arising from vacuum birefringence in the spectropolarimetric data of astrophysical sources. We collect five blazars with multiwavelength polarization measurements in different optical bands $(UBVRI)$. Taking into account the observed polarization angle contributions from both the intrinsic polarization angle and the rotation angle induced by LIV, and assuming that the intrinsic polarization angle is an unknown constant, we obtain new constraints on LIV by directly fitting the multiwavelength polarimetric data of the five blazars. Here we show that the birefringence parameter $\eta$ quantifying the broken degree of Lorentz invariance is limited to be in the range of $-8.91\times 10^{-7}$ $\textless$ $\eta$ $\textless$ $2.93\times10^{-5}$ at the $2\sigma$ confidence level, which is as good as or represents one order of magnitude improvement over the results previously obtained from ultraviolet/optical polarization observations. Much stronger limits can be obtained by future multiwavelength observations in the gamma-ray energy band.
}
\begin{document}
%%%%%%%%%%%%%%%%%%%%%%%%%%%%%%%%%%%%%%%%%%

%%%%%%%%%%%%%%%%%%%%%%%%%%%%%%%%%%%%%%%%%%

\section{Introduction}

Lorentz invariance, which says that the relevant physical laws of a non-accelerated physical system remain invariant under Lorentz transformations, is a fundamental symmetry of Einstein's theory of relativity. However, many quantum gravity theories seeking to unify quantum mechanics and general relativity predict that Lorentz invariance may be broken at the Planck energy scale $E_\mathrm{pl}\approx1.22\times10^{19}$ GeV \cite{ref:Kostelecky1989,ref:Kostelecky1991,ref:Kostelecky1995,ref:Mitrofanov2003,ref:Mattingly2005,
ref:Amelino-Camelia1998,ref:Tasson2014,ref:Amelino-Camelia2013}. Although these high energies are unattainable experimentally,
tiny deviations from Lorentz invariance may still exist at lower energies, motivating sensitive tests of Lorentz invariance.

In the photon sector, vacuum dispersion and vacuum birefringence are both significant characteristics of Lorentz invariance violation (LIV). Although these effects are believed to be very tiny at energies $\ll E_\mathrm{pl}$, they can become detectable by accumulating over large distances. Vacuum dispersion causes arrival-time differences of photons with different energies emitted simultaneously from an astrophysical source. Many attempts to set constraints on LIV have been performed based on
time-of-flight tests of high energy photons \cite{ref:Amelino-Camelia1998,ref:Biller1999,ref:Kaaret1999,ref:Kostelecky2008,ref:Pavlopoulos2005,ref:Ellis2006,ref:Jacob2008,ref:Kostelecky2009,
ref:Abdo2009a,ref:Abdo2009b,ref:Chang2012,ref:Nemiroff2012,ref:Vasileiou2013,ref:Ellis2013,ref:Kislat2015,ref:Zhang2015,ref:Wei2017c,
ref:Wei2017b,ref:Wei2017a,ref:Ellis2019,ref:Acciari2020}. In some specific cases, especially in loop quantum gravity, the deviations from Lorentz invariance take a particular form in which photons with right- and left-handed polarization propagate at different velocities. This leads to an energy-dependent rotation of the polarization vector of the linearly polarized photons, known as vacuum birefringence. Hence, Lorentz invariance can also be tested with astrophysical polarization measurements \cite{ref:Wei2020,ref:Wei2019b,ref:Friedman2019,ref:Kislat2017,ref:Lin2016,ref:Gotz2014,ref:Gotz2013,ref:Toma2012,ref:Stecker2011,ref:Laurent2011,
ref:Gubitosi2009,ref:Fan2007,ref:Jacobson2004,ref:Mitrofanov2003,ref:Kostelecky2013,ref:Kostelecky2007,ref:Kostelecky2006,ref:Kostelecky2001,
ref:Gleiser2001,ref:Colladay1998,ref:Carroll1990}.
As polarization measurements are more sensitive than time-of-flight measurements by a factor $\propto 1/E$, where $E$ is the energy of the light, more stringent constraints on LIV result from the former rather than the latter \cite{ref:Kostelecky2009}.
However, some Lorentz-violating theories do not predict any vacuum birefringence, so limits from time-of-flight measurements are
most interesting for investigating nonbirefringent effects.

In the literature, the linear polarization measurements have been applied to set limits on the birefringence parameter $\eta$.
The presence of linear polarization in the $\gamma$-ray band of gamma-ray bursts (GRBs) provides the current best limits on $\eta$,
i.e., $\eta < \mathcal{O}(10^{-16})$ \cite{ref:Wei2019b,ref:Lin2016,ref:Gotz2014,ref:Gotz2013,ref:Toma2012}. These upper limits
are obtained based on the indirect argument that vacuum birefringence would significantly deplete the net polarization of
the signal over a broad bandwidth. In addition, Fan et al. \cite{ref:Fan2007} used multiwavelength polarimetric data in order to
directly search for a frequency-dependent change of the linear polarization angle resulting from vacuum birefringence. They obtained a combined limit of $-2\times10^{-7} < \eta < 1.4\times10^{-7}$ by fitting the multiwavelength polarization observations of the optical afterglows of GRB 020813 and GRB 021004. However, it should be noted that the number of available linear polarization measurements in the $\gamma$-ray band for this particular test is very limited. Therefore, the outcomes of LIV tests lack significant statistical robustness even though some upper limits of $\eta$ are extraordinarily small.
It is fortunate that some astrophysical sources with multiwavelength polarization measurements in the optical band have been detected. Using different multiwavelength polarimetric data to constrain LIV is always helpful.

In this work, we gather 37 groups of multiwavelength polarization observations from five blazars to test LIV through the method that was discussed in Fan et al. \cite{ref:Fan2007}. The contents of this paper are organized as follows. In \S\ref{sec:Model}, we describe the method of constraining LIV. The multiwavelength polarization samples at our disposal and the resulting constraints on LIV are shown in \S\ref{sec:Results}. The physical implications of our results
are discussed in \S\ref{sec:Discuss}. Finally, our conclusions are presented in \S\ref{sec:Conclude}.

%%%%%%%%%%%%%%%%%%%%%%%%%%%%%%%%%%%%%%%%%%
\section{Method of constraining LIV}\label{sec:Model}

Considering Lorentz-violating effects, photons with right- and left- circular polarizations will traverse with different group velocities, which can be described using the leading term of a Taylor series expansion in the form \cite{ref:Mitrofanov2003}
\begin{equation}\label{eq:(1)}
v_{\pm} = c\left[1\pm\eta \left(\frac{\hbar\omega}{E_\mathrm{pl}}\right)^{n}\right],
\end{equation}
where $E_\mathrm{pl}\approx 1.22\times10^{19}$ GeV is the Planck energy, the $n$th-order expansion of the leading term stands for linear ($n=1$) or quadratic ($n=2$) energy dependence, $\pm$ represents the different polarization states for photons, and $\eta$ is a dimensionless constant characterizing the degree of the LIV effect. Here we consider the linearly polarized light from some astronomical sources, which is a superposition of two monochromatic waves with opposite circular polarizations. If Lorentz invariance is broken (i.e., when $\eta \neq 0$), then photons with right- and left-handed circular
polarizations emitted simultaneously from the same source should have different group velocities, leading to a rotation of the polarization vector of a linearly
polarized plane. The rotation angle in the differential propagation distance $dL_{(z)}$ can be expressed as \cite{ref:Mitrofanov2003,ref:Jacobson2004,ref:Gleiser2001,ref:Gambini1999}
\begin{equation}\label{eq:(2)}
\mathrm{d}\phi=\eta\left(\frac{\omega l_{p}}{c}\right)^{n+1} \frac{\mathrm{d}L_{(z)}}{l_{p}},
\end{equation}
where $l_{p} = \sqrt{\hbar G/c^{3}}=\hbar c/E_\mathrm{pl}$ is the Planck length scale.
Taking account of the cosmological expansion, the rotation angle $\Delta\phi_\mathrm{LIV}$ during the propagation from the source at redshift $z$ to the observer is given by
\begin{equation}\label{eq:(3)}
\Delta\phi_\mathrm{LIV}= \frac{\eta}{ H_{0}}\left(\frac{l_{p}}{c}\right)^{n}\left(2\pi\nu\right)^{n+1}F(z,n),
\end{equation}
where $\nu$ is the frequency of the observed light and $H_{0}$ is the Hubble constant. The function $F(z,n)$ depends on the cosmology model.
For a flat $\Lambda$CDM model,
\begin{equation}\label{eq:(4)}
F(z,n)=\int_{0}^{z}\frac{(1+z')^{n}dz'}{\sqrt {\Omega_M\left(1+z'\right)^3+\Omega_\Lambda} },
\end{equation}
where $\Omega_M$ and $\Omega_\Lambda$ are the cosmological parameters. Since the polarization data used in this work are not sensitive to higher-order terms, here
we only consider the linear ($n=1$) LIV case \cite{ref:Fan2007}. We thus have \cite{ref:Toma2012,ref:Laurent2011}
\begin{equation}\label{eq:(5)}
\Delta\phi_\mathrm{LIV}(E)= \eta \frac{E^2}{\hbar E_\mathrm{pl}H_{0}}\int_0^{z}\frac{(1+z')dz'}{\sqrt{\Omega_M(1+z')^3+\Omega_\Lambda}}.
\end{equation}
Throughout this paper, we adopt the cosmological parameters $H_0 = 67.36$ km $\rm s^{-1}$ $\rm Mpc^{-1}$, $\Omega_{M}=0.315$, and $\Omega_{\Lambda}=1-\Omega_{M}=0.685$ \cite{ref:Planck2020}.

When considering both the intrinsic polarization angle $\phi_0$ and the rotation angle $\Delta \phi_\mathrm{LIV}$ caused by LIV simultaneously, the observed polarization angle at a certain energy $E$ emitted from an astronomical event should consist of two terms
\begin{equation}\label{eq:(6)}
\phi_\mathrm{obs}(E)=\phi_0+B E^2,
\end{equation}
where the slope $B = \eta\frac{ F(z)}{\hbar E_\mathrm{pl} H_{0}}$ denotes the contribution from LIV.
However, since the emission mechanism of astronomical sources is still poorly understood, it is difficult to distinguish
an intrinsic polarization angle at the source from a rotation angle of the polarization plane induced by LIV.
Following Fan et al. \cite{ref:Fan2007}, we simply assume that all photons in the observed bandpass are radiated with
the same (unknown) intrinsic polarization angle. This assumption is vulnerable to unknown systematic uncertainties
associated with the unknown intrinsic properties of any given sources. In principle, these unknown intrinsic systematic
uncertainties can be minimized by analyzing the sizeable multiwavelength polarization data of a given source. However,
there are no abundant spectropolarimetry data in the current observations. We explore here the implications and limits
that can be set under the simplest assumption that the intrinsic polarization angle is an unknown constant. In this case, if the corresponding $E^2$ is taken as an independent variable, the free parameters $\phi_0$ and $B$ can be optimized from the linear fitting of the linear polarization measurements in several bands from a given astronomical event. With the fitting results, we then derive the limited range of the birefringence parameter $\eta$.
All the results are obtained at the 2$\sigma$ confidence level.

\section{Tests of LIV with polarized sources}\label{sec:Results}

It is obvious from Equation~(\ref{eq:(5)}) that the higher the energy band of the polarization observation and the larger the distance from the source, the better the constraint is on the birefringence parameter $\eta$. However, there are no multiwavelength polarization measurements in the
$\gamma$-ray or X-ray energy band, yet. Fan et al. \cite{ref:Fan2007} explored the limits of $\eta$ by observations of multiwavelength polarization from the optical afterglows of GRB 020813 and GRB 021004 (see also Ref. \cite{ref:Wei2020}).
Here we continue to search for astrophysical sources with multiwavelength polarimetric data that are suitable for testing LIV effects. The optical spectropolarimetry from several well-known blazars have been reported. We collect the high-quality polarimetry data of blazars 3C 66A, S5 0716+714, OJ 287, MK 421, and PKS 2155-304 from Refs. \cite{ref:Tommasi2001,ref:Takalo1994,ref:Takalo1993}. We then apply these polarized blazars to constrain possible birefringence effects in this work.

The linear polarization observations of these five blazars are presented in Table \ref{tab:observation}, including
the following information for each object: the energy bands in which polarization is observed; the observed linear polarization angles; and the corresponding uncertainties of the polarization angles.
These polarization observations have been obtained simultaneously in five optical bands ($UBVRI$) for all blazars.
The redshifts of the five blazars 3C 66A, S5 0716+714, OJ 287, MK 421, and PKS 2155-304 are 0.444, 0.310, 0.306, 0.031, and 0.116, respectively.
Since the optical ($UBVRI$) polarization measurements are carried out at different time periods,
there are several groups of spectropolarimetry data for each blazar. A total of 37 groups of multiwavelength polarimetric data
of the five blazars are listed in Table \ref{tab:observation}.
These enable us to carefully investigate the long- and short-term polarization variability of each blazar and to search for LIV effects.

We compile the polarization data from these bright blazars, and get the wavelengths ($\lambda_{U_\mathrm{eff}} = 0.360$ $\mu$m, $\lambda_{B_\mathrm{eff}} = 0.440$ $\mu$m, $\lambda_{V_\mathrm{eff}} = 0.530$ $\mu$m, $\lambda_{R_\mathrm{eff}} = 0.690$ $\mu$m and $\lambda_{I_\mathrm{eff}} = 0.830$ $\mu$m) as well as the corresponding polarization angles.
If each group of spectropolarimetry data is separately used to constrain the parameter $\eta$,
we would have slightly different results for the same blazar. Therefore, we allow a variation of the polarization from one time
period to another and consider only the time-averaged polarization angles for each blazar.
We take blazar 3C 66A as an example, which has eight groups of polarization data. For every wavelength bin, the time-averaged polarization angles for 3C 66A during eight observational periods are computed through $\overline{\phi}_{\rm obs}=\sum_{i}\phi_{{\rm obs},i}/8$.
The scattering in the shift between $\phi_{{\rm obs},i}$ and $\overline{\phi}_{\rm obs}$, i.e.,
$\sigma_{\phi_{i}}=|\phi_{{\rm obs},i}-\overline{\phi}_{\rm obs}|$, gives an estimate of the errors in $\overline{\phi}_{\rm obs}$,
i.e., $\sigma_{\overline{\phi}}=(\sum_{i}\sigma^{2}_{\phi_{i}})^{1/2}/8$.

\begin{figure}[H]
\centering
\includegraphics[width=7.75cm]{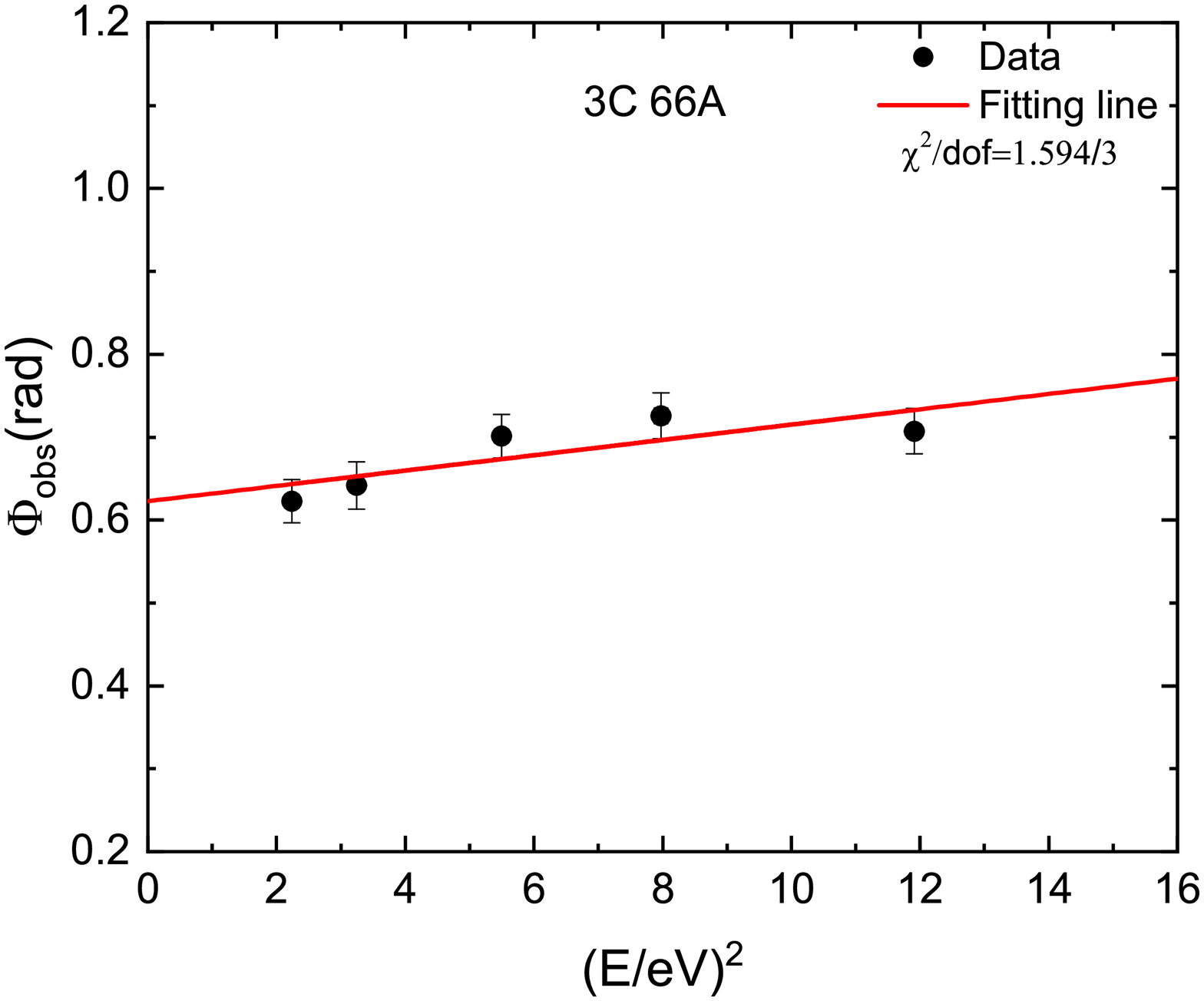}
\includegraphics[width=7.75cm]{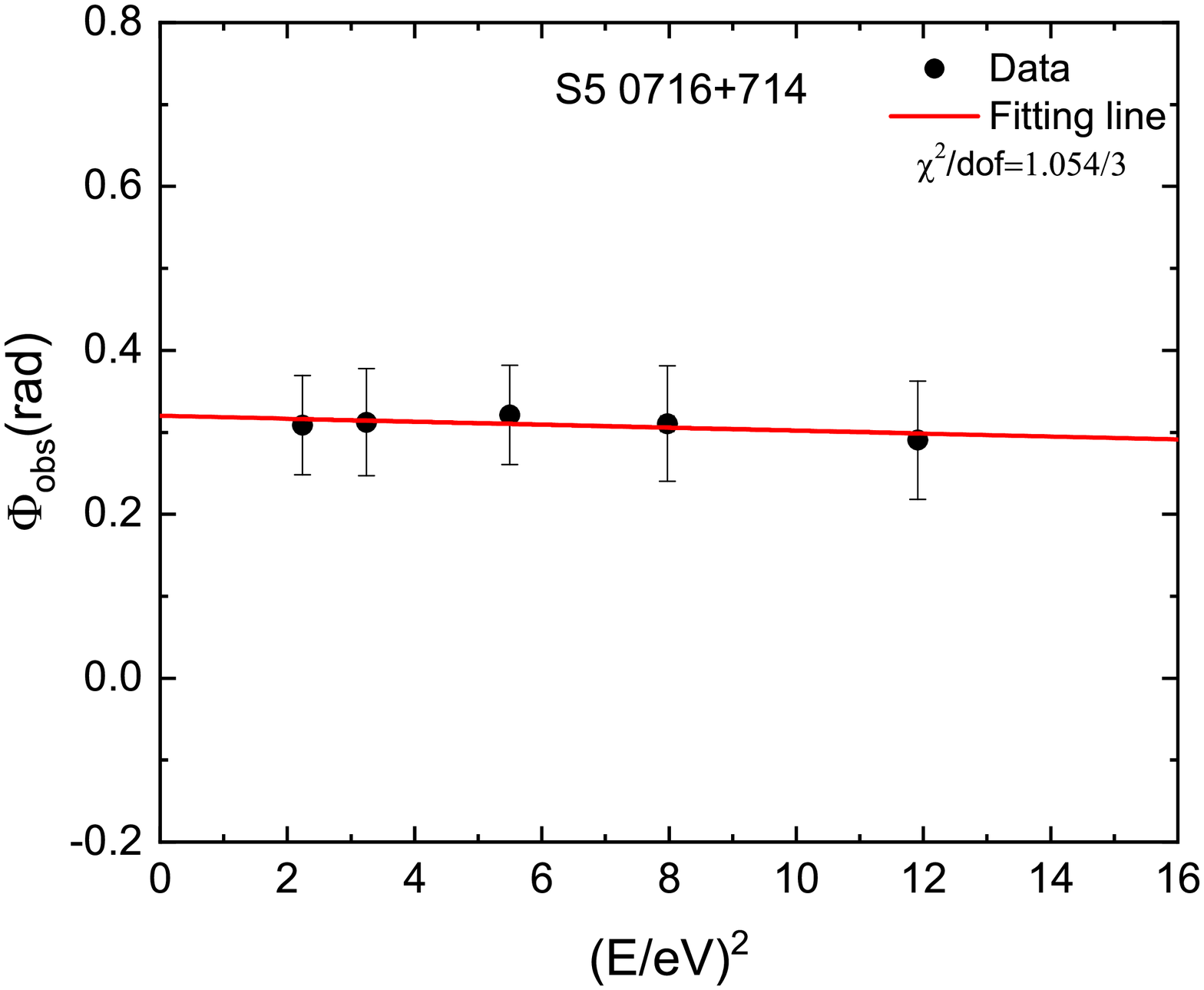}
\includegraphics[width=7.75cm]{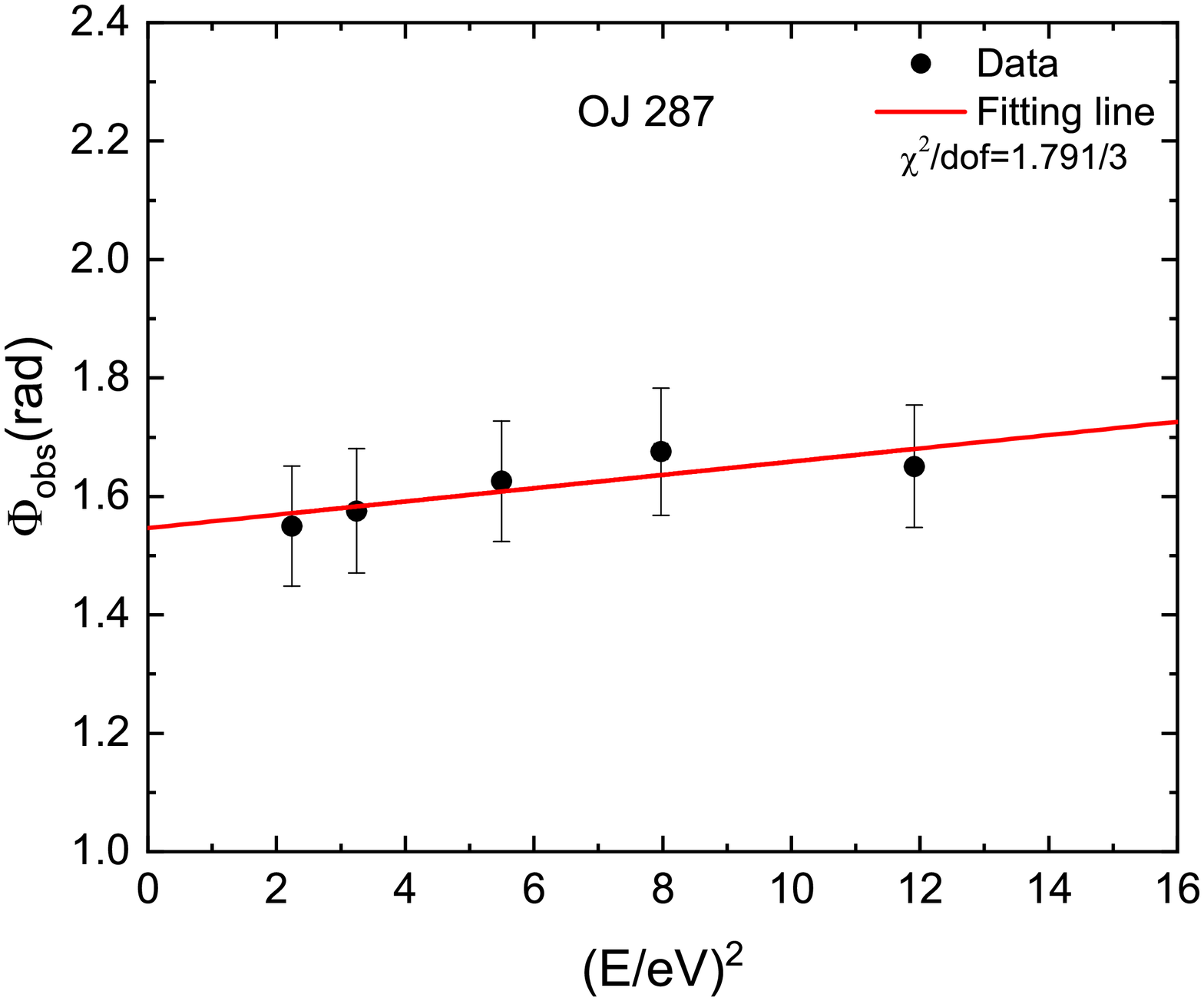}
\includegraphics[width=7.75cm]{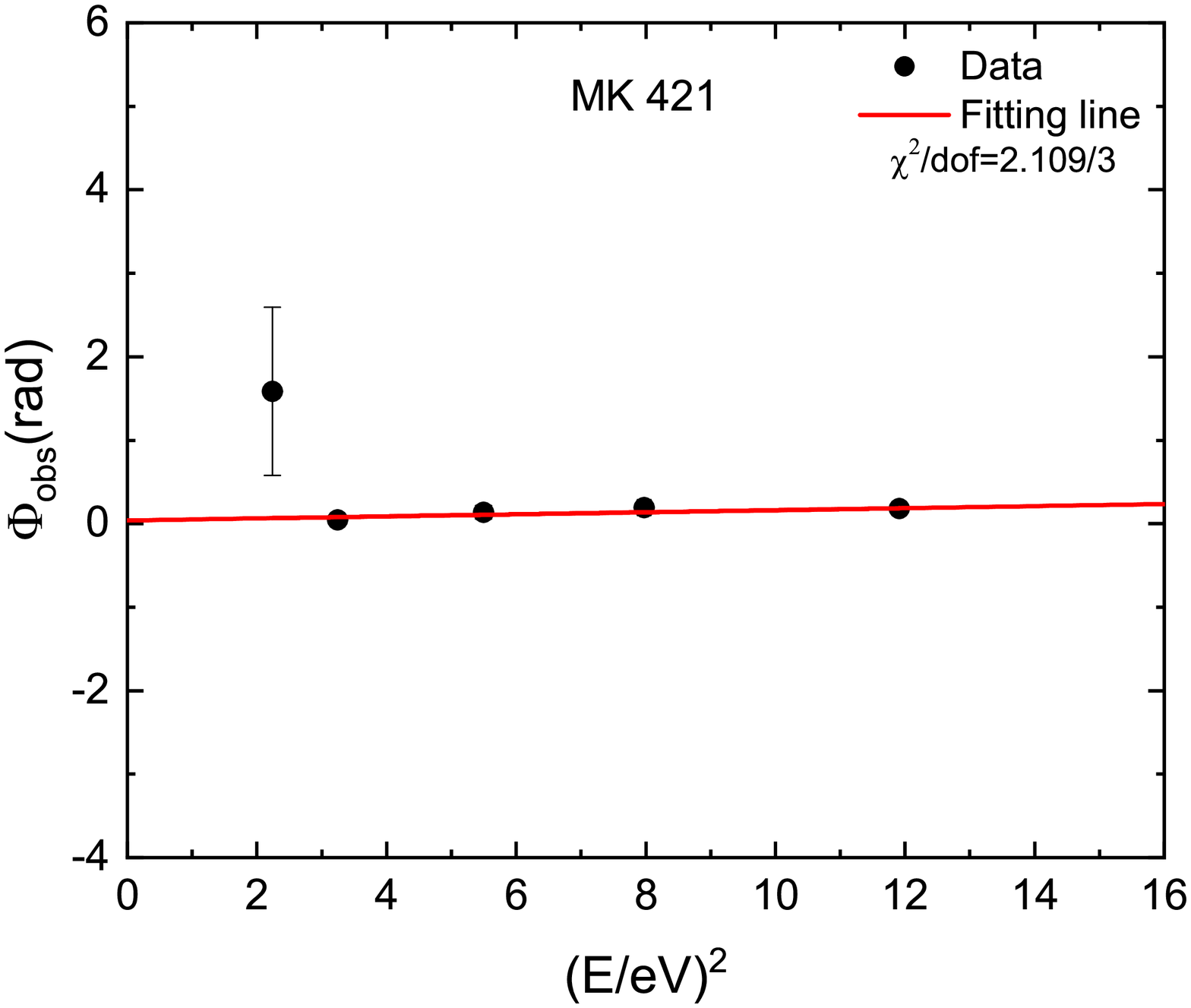}
\includegraphics[width=7.75cm]{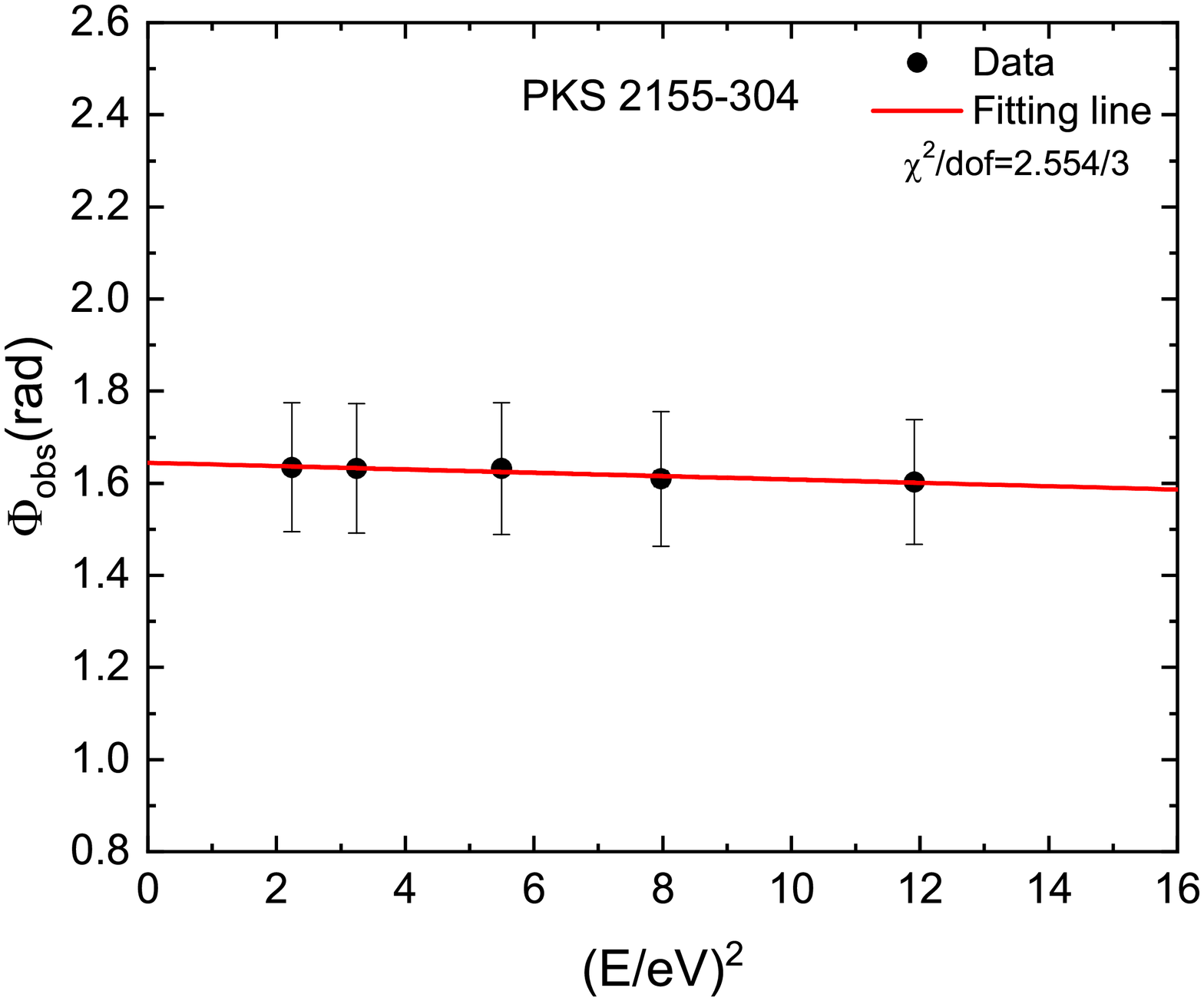}

\caption{Best-fitting results of the optical spectropolarimetric data of five blazars. The black dots correspond to the time-averaged linear-polarization data, and the red lines are the theoretical curves based on Equation (\ref{eq:(6)}).}
\label{fig:fit}
\end{figure}

We now have five groups of polarization data for the selected five blazars compiled with the time-averaged polarization angles and their corresponding energies.
For the linear regression applied to the multiwavelength polarization observations of each blazar, we use the function ${\phi}_{\rm obs}$ (Equation (\ref{eq:(6)})) to fit the observed data. The Metropolis-Hastings Markov Chain Monte Carlo (MHMC) method is applied to simulate the free parameters ${\phi}_{0}$ and $B$ of the linear function. The best-fitting results of
multiwavelength polarization observations for five blazars are presented in Table \ref{tab:results}. For each blazar in turn, the best-fitting values and $2\sigma$ uncertainties are provided for the parameters ${\phi}_{0}$ and $B$ and the corresponding
birefringence parameter $\eta$.
The time-averaged polarization angles of the five blazars as a function of $E^2$ are shown in Figure \ref{fig:fit}, and the red lines are the best-fitting lines. As shown in Figure \ref{fig:fit}, the polarization angles of blazars S5 0716+714 and PKS 2155-304 are proportional to $E^2$ with a negative coefficient, and the best-fitting values are $\eta=(-9.63\pm4.640)\times10^{-8}$ ($\phi_{0}=0.320\pm0.058$ rad) and $\eta=(-5.32\pm3.587)\times10^{-7}$ ($\phi_{0}=1.645\pm0.018$ rad), respectively. While the rest of the blazars show a positive coefficient with the fitting parameters $\eta=(3.386\pm1.249)\times10^{-7}$ ($\phi_{0}=0.623\pm0.024$ rad), $\eta=(6.546\pm22.769)\times10^{-6}$ ($\phi_{0}=0.046\pm0.482$ rad), and $\eta=(6.226\pm7.715)\times10^{-7}$ ($\phi_{0}=1.544\pm0.094$ rad) for 3C 66A, MK 421, and OJ 287, respectively. Our constraints show that except for 3C 66A at the $5.4\sigma$ confidence level and PKS 2155-304 at the $3.0\sigma$ confidence level, the data set of the other three blazars are consistent with the possibility of no LIV at all
(i.e., $\eta = 0$) within the $1.8\sigma$ confidence level. The wavelength dependence of polarization angles, even if
present in 3C 66A and PKS 2155-304, cannot be due to LIV effects as this would clash with the best limit from S5 0716+714 by factors of 4 to 6, and must therefore be of intrinsic astrophysical origin.
At the $2\sigma$ confidence level, the limits on $\eta$ from five blazars are $-8.91\times 10^{-7} < \eta <  2.93\times10^{-5}$.

\section{Discussion}\label{sec:Discuss}

According to the summary of the limits on the birefringence parameter $\eta$ from the polarization measurements of various astrophysical sources \cite{ref:Wei2021}, we can simply make a comparison with our results. The polarization observations on the prompt $\gamma$-ray emission of GRBs place the hitherto most stringent constraints on $\eta$, i.e., $\eta < \mathcal{O}(10^{-16})$ \cite{ref:Wei2019b,ref:Lin2016,ref:Gotz2014,ref:Gotz2013,ref:Toma2012}. As expected, our optical polarization constraints are not competitive with that from the observations of gamma-ray polarization. However, these gamma-ray polarization
limits suffer from a relatively low statistical confidence of the polarization measurements and large systematic uncertainties. In the optical band as we do, Gleiser \& Kozameh \cite{ref:Gleiser2001} set an upper limit of $\eta < 10^{-4}$ by processing the linearly polarized ultraviolet light from the radio galaxy 3C 256, and Fan et al. \cite{ref:Fan2007} obtained the limits on $\eta$ of $-2 \times 10^{-7} < \eta < 1.4 \times 10^{-7}$ by using the ultraviolet/optical polarization data of the afterglows of GRBs.
We have used the optical polarimetric observations in five bands ($UBVRI$) of blazars to derive the limits $-8.91\times 10^{-7} < \eta <  2.93\times10^{-5}$ at the $2\sigma$ confidence level, which is essentially as good as or one order of magnitude improvement over previous limits obtained by ultraviolet/optical polarization observations.

It is worth pointing out that Kislat \& Krawczynski \cite{ref:Kislat2017} also used the optical polarization measurements to constrain LIV. They introduced Lorentz and CPT symmetry-violating effects by adding new terms which can be ordered by the mass dimension of the corresponding field operator in the Standard Model Lagrangian. They analyzed optical polarization data from 72 AGNs and GRBs and obtained a set of constraints on the 16 coefficients of mass dimension $d = 5$ of the Standard-Model Extension photon sector. Their results implied a lower limit on the energy scale of quantum gravity of $10^5$ times the Planck energy.
The approach in \cite{ref:Kislat2017} constrained LIV by limiting the coefficients of $d=5$, while we got the results by  constraining the birefringence parameter $\eta$.
Both approaches are rigorous and constrain LIV well.

Note that we have used the standard flat $\Lambda$CDM model for our calculations (see Eq. (\ref{eq:(4)})).
However, other different cosmological models have been proposed.
In these models, dark energy evolves in a way that is different from that in the $\Lambda$CDM model.
Therefore, it is necessary to study whether LIV is caused by the particular model of the universe we are applying.
Moreover, apart from dark energy, the contribution of spatial curvature should not be ignored.
Refs. \cite{ref:Biesiada2009} and \cite{ref:Luongo2012} have investigated these issues in different cosmological models.
By extending the analysis performed in the $\Lambda$CDM model to other cosmological models, they found that
the effects, e.g, the redshift dependence of time delays caused by LIV, also occur in alternative cosmological models.

It should be stressed that we have performed the analysis based on the assumption of isotropy. On the other hand, anisotropic LIV has been discussed
in theory, and constraints on anisotropic LIV have also been obtained from some relevant observations, see, e.g., Refs. \cite{ref:Wei2017b, ref:Kislat2015, ref:Kislat2017}.

%%%%%%%%%%%%%%%%%%%%%%%%%%%%%%%%%%%%%%%%%%
\section{Conclusion}\label{sec:Conclude}

As a consequence of LIV, the group velocities of left- and right-handed circularly polarized photons that are emitted from the same astrophysical source should differ slightly, leading to vacuum birefringence and a frequency-dependent rotation of the polarization vector of linearly polarized light. LIV can therefore be tested with linear polarization measurements of astrophysical sources. A key challenge in the idea of searching for a frequency-dependent change in polarization, however,
is to distinguish the rotation angle induced by LIV from any source-intrinsic polarization angle in the emission of photons
in different energy bands. Following the method of Fan et al. \cite{ref:Fan2007}, we simply assume that the intrinsic polarization angle is an unknown constant and expect to observe birefringence effects as a frequency-dependent linear polarization vector.

To constrain LIV, we try to search for a similar frequency-dependent trend in multiwavelength polarization observations of
astrophysical sources. In this work, five blazars with multiwavelength polarization angles in five optical bands ($UBVRI$) are collected. For each blazar, the optical polarization measurements were carried out at different time periods, thus
there are a total of 37 groups of spectropolarimetry data for the five blazars.
Considering the temporal variation of the polarization observations, we extract the time-averaged polarization angles in five wavelength bins for each selected blazar. We then use the function ${\phi}_{\rm obs}(E)$ (Equation (\ref{eq:(6)})) to fit the time-averaged polarization angles versus the quantities of $E^{2}$, and obtain the best-fitting results of the intrinsic polarization angle $\phi_{0}$ and the birefringence parameter $\eta$.
At the $2\sigma$ confidence level, we find that $\eta$ is constrained to be in the range of $-8.91\times 10^{-7} < \eta <  2.93\times10^{-5}$ for five blazars.
These results are comparable with or represent sensitivities improved by one order
of magnitude over existing limits from ultraviolet/optical polarization observations. More stringent limits on the birefringence parameter can be expected as the analysis presented here is applied to a larger number of astrophysical sources with larger distances and higher energy polarimetry.

Note that the rotation of the linear polarization plane can also be caused by magnetized plasmas, where this phenomenon is known as Faraday rotation.
So it is necessary to investigate whether our resulting constraints could be affected by Faraday rotation. The dependence of the rotation angle on Faraday rotation is $\Delta\Phi_\mathrm{Far}\propto E^{-2}$, different from its dependence on LIV effects as
$\Delta\Phi_\mathrm{LIV}\propto E^{2}$ shown in Equation (\ref{eq:(5)}).
According to the discussions of Fan et al. \cite{ref:Fan2007} and Wei \& Wu \cite{ref:Wei2020}, the rotation angle $\Delta\Phi_\mathrm{Far}$ is extremely small at the optical and higher energy band. Therefore, the Faraday rotation can be ignored in our analysis.

%%%%%%%%%%%%%%%%%%%%%%%%%%%%%%%%%%%%%%%%%%
\vspace{6pt} 

%%%%%%%%%%%%%%%%%%%%%%%%%%%%%%%%%%%%%%%%%%
%% optional
%\supplementary{The following are available online at \linksupplementary{s1}, Figure S1: title, Table S1: title, Video S1: title.}


%%%%%%%%%%%%%%%%%%%%%%%%%%%%%%%%%%%%%%%%%%
% \authorcontributions{For research articles with several authors, a short paragraph specifying their individual contributions must be provided. The following statements should be used ``Conceptualization, X.X. and Y.Y.; methodology, X.X.; software, X.X.; validation, X.X., Y.Y. and Z.Z.; formal analysis, X.X.; investigation, X.X.; resources, X.X.; data curation, X.X.; writing--original draft preparation, X.X.; writing--review and editing, X.X.; visualization, X.X.; supervision, X.X.; project administration, X.X.; funding acquisition, Y.Y. All authors have read and agreed to the published version of the manuscript.'', please turn to the  \href{http://img.mdpi.org/data/contributor-role-instruction.pdf}{CRediT taxonomy} for the term explanation. Authorship must be limited to those who have contributed substantially to the work reported.}

%%%%%%%%%%%%%%%%%%%%%%%%%%%%%%%%%%%%%%%%%%

%%%%%%%%%%%%%%%%%%%%%%%%%%%%%%%%%%%%%%%%%%
\acknowledgments{We thank Yu-Peng Yang for helpful discussions. This work is supported by the National Natural Science Foundation
of China (Grant Nos. 11703015, U2038106, 11725314, U1831122, and 12041306), the Youth Innovations and Talents
Project of Shandong Provincial Colleges and Universities (Grant No. 201909118),
the Youth Innovation Promotion Association (2017366), the Key Research Program of Frontier Sciences (grant No.
ZDBS-LY-7014), the Strategic Priority Research Program
``Multi-waveband gravitational wave universe'' (grant No. XDB23000000) of the
Chinese Academy of Sciences, and the Major Science and Technology Project of Qinghai
Province (2019-ZJ-A10).}
%%%%%%%%%%%%%%%%%%%%%%%%%%%%%%%%%%%%%%%%%%
\conflictsofinterest{The authors declare no conflict of interest.}

 %%%%%%%%%%%%%%%%%%%%%%%%%%%%%%%%%%%%%%%%%%%%%%%
\clearpage

%\longtab{}{

\clearpage

\begin{table}[H]
\caption{Measurements of multiwavelength polarization for different blazars.}
\centering
%% \tablesize{} %% You can specify the fontsize here, e.g., \tablesize{\footnotesize}. If commented out \small will be used.
\begin{tabular}{cccc|cccccccc}
\toprule
\textbf{Group}	& \textbf{Filter}	& \textbf{Polarization angle\,$(^\circ)$} &   \textbf{$\sigma$} &
\textbf{Group}	& \textbf{Filter}	& \textbf{Polarization angle\,$(^\circ)$} &   \textbf{$\sigma$}\\
\midrule
                    &       &   \textbf{3C 66A} &     &      &     &   \textbf{OJ 287}  &       &\\
         1          &   U   &   40.5    &   0.7 &  1  &   U   &   108.6   &   1.8 &\\
                    &   B   &   43.1    &   0.6 &     &   B   &   108.1   &   2.5 &\\
                    &   V   &   40.1    &   0.6 &     &   V   &   108.7   &   1.1 &\\
                    &   R   &   35.1    &   0.5 &     &   R   &   103.7   &   1.6 &\\
                    &   I   &   32.2    &   1.2 &     &   I   &   98.4    &   3.3 &\\

         2          &   U   &   41.0    &   0.6 & 2   &   U   &   103.9   &   1.0 &\\
                    &   B   &   42.0    &   0.6 &     &   B   &   109.3   &   1.5 &\\
                    &   V   &   41.0    &   0.6 &     &   V   &   105.2   &   1.5 &\\
                    &   R   &   35.3    &   0.5 &     &   R   &   101.5   &   1.0 &\\
                    &   I   &   31.9    &   1.2 &     &   I   &   106.2   &   4.2 &\\

         3          &   U   &   39.3    &   0.5 & 3   &   U   &   105.2   &   1.3 &\\
                    &   B   &   40.2    &   0.8 &     &   B   &   106.4   &   1.7 &\\
                    &   V   &   38.1    &   0.8 &     &   V   &   106.3   &   1.6 &\\
                    &   R   &   33.8    &   0.4 &     &   R   &   101.4   &   1.2 &\\
                    &   I   &   35.3    &   1.7 &     &   I   &   97.6    &   1.9 &\\

         4          &   U   &   39.6    &   1.2 & 4   &   U   &   106.8   &   1.5 &\\
                    &   B   &   41.6    &   1.3 &     &   B   &   107.8   &   1.4 &\\
                    &   V   &   39.7    &   1.4 &     &   V   &   103.5   &   1.4 &\\
                    &   R   &   33.8    &   1.7 &     &   R   &   103.2   &   0.9 &\\
                    &   I   &   36.0    &   3.1 &     &   I   &   104.5   &   3.2 &\\

         5          &   U   &   51.3    &   0.2 & 5   &   U   &   105.7   &   1.3 &\\
                    &   B   &   51.7    &   0.2 &     &   B   &   107.4   &   2.1 &\\
                    &   V   &   50.4    &   0.2 &     &   V   &   102.1   &   1.4 &\\
                    &   R   &   48.1    &   0.2 &     &   R   &   102.3   &   0.8 &\\
                    &   I   &   45.7    &   0.5 &     &   I   &   96.8    &   3.2 &\\

         6          &   U   &   39.3    &   0.2 & 6   &   U   &   105.6   &   1.3 &\\
                    &   B   &   40.2    &   0.3 &     &   B   &   108.5   &   1.4 &\\
                    &   V   &   39.4    &   0.3 &     &   V   &   103.6   &   1.2 &\\
                    &   R   &   38.3    &   0.2 &     &   R   &   102.0   &   1.3 &\\
                    &   I   &   36.9    &   0.6 &     &   I   &   102.6   &   2.9 &\\

         7          &   U   &   38.2    &   0.2 & 7   &   U   &   108.2   &   1.8 &\\
                    &   B   &   39.1    &   0.2 &     &   B   &   109.8   &   2.8 &\\
                    &   V   &   38.2    &   0.3 &     &   V   &   104.0   &   2.1 &\\
                    &   R   &   37.2    &   0.2 &     &   R   &   101.0   &   1.7 &\\
                    &   I   &   35.7    &   0.4 &     &   I   &   99.8    &   3.1 &\\

         8          &   U   &   35.0    &   0.4 & 8   &   U   &   107.5   &   1.3 &\\
                    &   B   &   34.8    &   0.3 &     &   B   &   108.7   &   1.7 &\\
                    &   V   &   34.5    &   0.4 &     &   V   &   104.2   &   1.4 &\\
                    &   R   &   32.6    &   0.3 &     &   R   &   100.9   &   1.0 &\\
                    &   I   &   31.8    &   0.6 &     &   I   &   94.0    &   3.2 &\\

                    &       & \textbf{S5 0716+714}&   & 9     &   U   &   54.6    &   1.1 &\\
         1          &   U   &   22.5    &   0.4 &     &   B   &   53.4    &   1.2 &\\
                    &   B   &   23.5    &   0.3 &     &   V   &   51.7    &   1.6 &\\
                    &   V   &   23.3    &   0.3 &     &   R   &   44.1    &   1.1 &\\
                    &   R   &   23.2    &   0.3 &     &   I   &   44.8    &   2.8 &\\
                    &   I   &   22.6    &   0.5 & 10  &   U   &   65.5    &   0.7 &\\

         2          &   U   &   10.8    &   0.2 &     &   B   &   66.9    &   0.7 &\\
                    &   B   &   12.1    &   0.3 &     &   V   &   65.3    &   0.7 &\\
                    &   V   &   13.5    &   0.3 &     &   R   &   64.4    &   0.5 &\\
                    &   R   &   12.6    &   0.3 &     &   I   &   63.6    &   1.4 &\\
                    &   I   &   12.8    &   0.5 & 11  &   U   &   68.8    &   0.3 &\\

\bottomrule
\end{tabular}
\label{tab:observation}
\end{table}

\begin{table}[H]\ContinuedFloat
\caption{{\em Cont.}}
\centering
%% \tablesize{} %% You can specify the fontsize here, e.g., \tablesize{\footnotesize}. If commented out \small will be used.
\begin{tabular}{cccc|cccccc}
\toprule
\textbf{Group}	& \textbf{Filter}	& \textbf{Polarization angle\,$(^\circ)$} &   \textbf{$\sigma$} &
\textbf{Group}	& \textbf{Filter}	& \textbf{Polarization angle\,$(^\circ)$} &   \textbf{$\sigma$}\\
\midrule
                    &   B   &   69.7    &   0.6 &      &   V   &   119.4   &   2.6 &\\
                    &   V   &   70.0    &   0.4 &      &   R   &   120.1   &   1.8 &\\
                    &   R   &   68.4    &   0.5 &      &   I   &   120.0   &   7.7 &\\
                    &   I   &   68.5    &   1.5 &  7   &   U   &   68.9    &   4.4 &\\
                  \\
                    &       &   \textbf{MK 421}  &  &   &   B   &   68.2   &   3.2 &\\
        1           &   U   &   10.7    &   2.4 &      &   V   &   69.4    &   2.8 &\\
                    &   B   &   12.6    &   2.7 &      &   R   &   69.5    &   1.8 &\\
                    &   V   &   14.2    &   2.3 &      &   I   &   70.1    &   3.8 &\\
                    &   R   &    4.4    &   2.1 & 8    &   U   &   69.1    &   2.5 &\\
                    &   I   &    9.4    &   6.0 &      &   B   &   70.3    &   3.6 &\\

        2           &   U   &   10.5    &   4.0 &      &   V   &   72.4    &   2.1 &\\
                    &   B   &    9.3    &   2.9 &      &   R   &   72.5    &   1.7 &\\
                    &   V   &    1.5    &   2.7 &      &   I   &   73.7    &   3.3 &\\
                    &   R   &    0.6    &   1.3 & 9    &   U   &   63.1    &   2.2 &\\
                    &   I   &   172.6   &   2.6 &      &   B   &   64.5    &   1.9 &\\
                    \\
                    &       &\textbf{PKS 2155-304}& &   &   V   &   66.7   &   1.6 &\\
        1            &   U   &   93.1    &   3.9 &     &   R   &   67.0    &   1.3 &\\
                     &   B   &   96.4    &   2.9 &     &   I   &   67.7    &   2.5 &\\
                     &   V   &   96.7    &   2.7 & 10  &   U   &   113.8    &   7.3 &\\
                     &   R   &   97.1    &   2.7 &     &   B   &   113.7    &   5.7 &\\
                     &   I   &   96.5    &   4.6 &     &   V   &   116.6    &   4.9 &\\

        2            &   U   &   12.0    &   5.8 &     &   R   &   115.9    &   4.7 &\\
                     &   B   &   4.3     &   7.4 &     &   I   &   117.1    &   5.3 &\\
                     &   V   &   6.4     &   3.8 & 11  &   U   &   105.0    &   2.6 &\\
                     &   R   &   8.6     &   3.7 &     &   B   &   105.7    &   2.5 &\\
                     &   I   &   8.3     &   11.3&     &   V   &   105.6    &   1.8 &\\

        3            &   U   &   101.3   &   3.0 &     &   R   &   106.1    &   1.3 &\\
                     &   B   &   104.1   &   3.9 &     &   I   &   106.0    &   3.0 &\\
                     &   V   &   102.4   &   2.0 & 12  &   U   &   109.8    &   3.5 &\\
                     &   R   &   101.5   &   2.3 &     &   B   &   110.4    &   4.3 &\\
                     &   I   &   101.0   &   3.4 &     &   V   &   109.3    &   2.4 &\\

        4            &   U   &   118.8   &   6.0 &     &   R   &   108.1    &   2.1 &\\
                     &   B   &   121.6   &   3.1 &     &   I   &   107.8    &   5.8 &\\
                     &   V   &   120.9   &   2.9 & 13  &   U   &   97.5    &   3.1 &\\
                     &   R   &   120.5   &   3.4 &     &   B   &   88.3    &   4.8 &\\
                     &   I   &   121.1   &   5.9 &     &   V   &   96.4    &   2.8 &\\

        5            &   U   &   125.4   &   4.0 &     &   R   &   95.5    &   2.5 &\\
                     &   B   &   130.1   &   3.2 &     &   I   &   99.3    &   4.3 &\\
                     &   V   &   131.5   &   1.6 & 14  &   U   &   91.8    &   3.8 &\\
                     &   R   &   131.3   &   3.0 &     &   B   &   94.3    &   5.3 &\\
                     &   I   &   129.6   &   4.7 &     &   V   &   95.3    &   8.5 &\\

       6             &   U   &   115.9   &   4.6 &     &   R   &   95.6    &   5.6 &\\
                     &   B   &   119.1   &   3.3 &     &   I   &   92.9    &   6.9 &\\

\bottomrule
\end{tabular}
\label{tab:observation}
\end{table}

\begin{table}[H]
\caption{Best-fitting results of multiwavelength polarization observations for five blazars.}
\centering
%% \tablesize{} %% You can specify the fontsize here, e.g., \tablesize{\footnotesize}. If commented out \small will be used.
\begin{tabular}{ccccccc}
\toprule
\textbf{Source}	& \textbf{\emph{z}}	& \textbf{B} & \textbf{$\phi_0$(rad)}  &  \textbf{$\eta(\times10^{-7})$}  \\
\midrule
3C 66A       &  0.444     &	  0.009$\pm$0.003	       &	0.623$\pm$0.024       &    3.386$\pm$1.249      &\\

S5 0716+714	 &  0.310     &	  -0.002$\pm$0.009	       &	0.320$\pm$0.058       &    -0.963$\pm$4.640     &\\

OJ 287       &  0.306 	  &   0.012$\pm$0.013          &    1.544$\pm$0.094       &    6.226$\pm$7.175     &\\

MK 421       &  0.031 	  &   0.012$\pm$0.041          &    0.046$\pm$0.482       &    65.464$\pm$227.692      &\\

PKS 2155-304 &  0.116     &   -0.004$\pm$0.002         &    1.645$\pm$0.018       &    -5.322$\pm$3.587      &\\

\bottomrule
\end{tabular}
\label{tab:results}
\end{table}

%%%%%%%%%%%%%%%%%%%%%%%%%%%%%%%%%%%%%%%%%%
\reftitle{References}

\end{document}